# Battery Swapping Station as an Energy Storage for Capturing Distribution-Integrated Solar Variability

Zohreh S. Hosseini, Mohsen Mahoor, and Amin Khodaei
Dept. of Electrical and Computer Engineering
University of Denver
Denver, CO, USA
Zohreh.Hosseini@du.edu, Mohsen.Mahoor@du.edu, Amin.Khodaei@du.edu

*Abstract*—Managing the inherent variability of solar generation is a critical challenge for utility grid operators, particularly as the distribution grid-integrated solar generation is making fast inroads in power systems. This paper proposes to leverage Battery Swapping Station (BSS) as an energy storage for mitigating solar photovoltaic (PV) output fluctuations. Using mixed-integer programming, a model for the BSS optimal scheduling is proposed to capture solar generation variability. The proposed model aims at minimizing the BSS total operation cost, which represents the accumulated cost of exchanging power with the utility grid. The model is subject to four sets of constraints associated with the utility grid, the BSS system, individual batteries, and solar variability. Numerical simulations on a test BSS demonstrate the effectiveness of the proposed model and show its viability in helping the utility grids host a higher penetration of solar generation.

## Nomenclature

*Indices and Sets:*
$b$    Index for battery.
$t$    Index for time.
$j$    Index for prosumers.
ch    Superscript for battery charging mode.
dch    Superscript for battery discharging mode.
N    Set of consumers and prosumers.

*Parameters:*
$D$    Battery swapping demand.
$\rho$    Electricity price.
$\tau$    Time duration.
$\eta^{ch}$    Charging efficiency.
$\eta^{dch}$    Discharging efficiency.
$M$    Large positive constant.
$\varepsilon$    Small positive constant.
$\Delta^u$    The variability limit provided by the utility grid.
$P^c$    Aggregated prosumers net load.

*Variables:*
$C$    Battery stored energy.
$P^{ch}$    Battery charging power.
$P^{dch}$    Battery discharging power.
$P^M$    Imported/exported power from/to the utility grid.
$P^u$    Distribution feeder net load supplied by the utility grid.
$x^F$    Battery fully-charged state (1 when fully-charged, 0 otherwise).

## I. Introduction

THE GLOBAL environmental concern regarding the use of fossil fuels in electricity generation has motivated many countries in deploying higher levels of renewable energy resources. Among renewable energy resources, solar photovoltaic (PV) is envisioned to be a major player in future power systems and a viable enabler of sustainable power generation. Solar energy is clean, widely available, and relatively low maintenance. Moreover, unlike traditional power generation resources, which are installed in a centralized manner, solar energy resources can be easily deployed as a distributed generation resource [1]-[4]. Solar energy resources have attracted consumers who are willing to make up part of their electricity consumption or even economically benefit from a local power generation [5],[6]. The dropping cost of solar technology and the state and governmental incentives have made the path for a rapid growth of solar generation. More than 7 GW of solar PV was installed in the U.S. in 2016, where residential PV with over 2 GW represented the biggest segment [7]. All in all, the solar generation is making fast inroads in power systems [8]-[10].

Although various methods are carried out in the literature for solar forecasting problem [11]-[14], they mainly suffer from a degree of inaccuracy due to inherent variability (i.e., intermittency and volatility) and uncertainty in solar generation. The intermittency indicates that the solar generation is not always available, while the volatility denotes the fluctuations of solar generation in different time scales such as seconds, minutes, and hours. Uncertainty indicates the failure of accurate forecast in the time and the magnitude of solar generation variability. These characteristics negatively impact the solar generation and necessitate the deployment of flexible energy resources to facilitate the integration of solar generation into power systems [15]-[17]. To this end, coordinating solar generation with battery energy storage systems is a common approach, where the coordinated scheme can pick up the variability of solar generation to achieve a smooth and controllable output power [18]-[21].

A novel and viable method for addressing the aforementioned challenges is to reap the benefit of available energy storage system in a Battery Swapping Station (BSS).

The idea of the BSS has been proposed to provide Electric Vehicle (EV) owner with a unique opportunity of exchanging an empty battery with a fully-charged one in designated stations. The distinguished features of this approach are that not only EVs can be charged in a short amount of time, but also the sticker price of the EVs will be reduced significantly as the battery cost is deducted from the total cost of the EV [22]. The BSS, as an effective alternative to existing EV charging methods such as Battery Charging Station (BCS), has attracted significant attention across the world including in China [23], India [24], and Germany [25], to name a few. Nevertheless, due to the rapid proliferation of EVs in automotive market, the BSS deployment is rapidly growing across the globe [26].

The concept of the BSS as an energy storage has been studied in the literature. Authors in [27] study a BSS-enabled power system with high penetration of renewable-based energy resources, where the BSS is utilized to provide fully-charged battery to EVs as well as to help with energy management. The optimal storage capacity of the BSS is obtained by analyzing the behavior of the power system with a high penetration of renewable energy resources. In [28], a study for evaluating the economic value of battery energy storage inside the BSS is proposed. The paper concludes that leveraging the batteries inside the BSS is more beneficial than pumped storage for managing surplus electricity generated by solar PV. The potential of providing regulation services by energy storage in BSS is investigated in [29] and [30]. Based on an interaction framework, called Station-to-Grid (S2G), the integration of BSS into power systems is presented. This framework is developed in a way that the BSS not only is in charge of battery swapping service for EVs, but also can offer regulation reserves. The simulation results carried out in the paper demonstrate that the BSS can mitigate frequency deviation as well as tie-line power fluctuations.

The primary objective of this paper is to provide a BSS-based framework to capture distribution grid-integrated solar variability. To this end, the BSS exchanged power with the utility grid is reshaped with the objective of mitigating distributed solar generation variability. A mixed-integer programming formulation is used for problem modeling.

The remainder of the paper is organized as follows. Section II discusses the BSS players and presents the BSS-based model outline for capturing distribution grid-integrated solar variability. The BSS optimal scheduling formulation is proposed in Section III. Section IV presents numerical analyses to show the effectiveness of the proposed model. Finally, conclusions are presented in Section V.

## II. Model Outline

### A. BSS Players

The main idea of introducing the BSS into the EV industry is that an EV owner can quickly swap an empty or a near-empty battery with a fully-charged one in a short time. To implement this innovative idea, at least three main players, including the EV owner, the BSS owner, and the power system, should take part. In what follows, the BSS will be investigated from each player's perspective.

From the EV owner's perspective, the BSS deployment has several benefits. The first and foremost advantage is that the sticker price of the EV is significantly reduced, as the battery is owned by the BSS instead of the EV owner. The battery replacement is processed faster than refuelling a gasoline-powered vehicle. By properly placing the BSS, the EV owner could easily plan for longer distance trips, and accordingly the range anxiety could be alleviated. Moreover, the EV owner is no longer concerned about the battery lifetime. As the BSS owner is in charge of upgrading charging facilities, such as high-power battery chargers, the relevant cost for upgrading household infrastructure is also entirely eliminated.

Power system operators are always looking for viable approaches to reduce network congestion and the peak load, for adding economic and reliability value to the grid by better utilizing available resources. However, the unpredictable behavior of EV owners for charging the batteries in plug-in mode could negatively impact power system operation in terms of increasing peak load and network congestion. The BSS approach could potentially turn this challenge into an opportunity by providing a scheduled charging/discharging strategy. By considering the aggregated batteries inside the BSS as a large shiftable load, the battery charging schedule could be shifted to the night time or off-peak hours in order to tackle the potential peak demand or overloading, caused by growing penetration of EVs. Nevertheless, the BSS can serve as a viable source of energy storage in the system to capture solar generation variability, as discussed in this paper.

The BSS owner could potentially achieve the largest benefits among players. Compared to the BCS, the BSS could minimize its electricity cost for battery charging/discharging by operating at a least-cost schedule based on hourly electricity prices. Furthermore, the BSS owner could make a profit via participating in electricity market and offering ancillary and reserve services to the grid. In terms of the cost of real estate, as the BSS owner does not need access to spacious parking lots, as compared to the BCS, substantial cost savings would be guaranteed. Eventually, since all batteries are unified and follow a consistent standard, as assumed in this paper, the battery charging/discharging process is convenient for the BSS owner.

### B. Optimal Scheduling Model

Consider a distribution network in which a BSS and several consumers with the ability of electricity generation, i.e., prosumers, are connected to a distribution feeder. The prosumers own distributed rooftop solar PV, where accordingly bring variability to the power required to be supplied by the utility grid. In addition, the behavior of prosumers for buying/selling electricity to/from the utility grid is uncontrolled, as they aim at maximizing benefits subject to their financial objectives (i.e., the minimum electricity payment). The BSS which is deployed at the distribution network not only aims at providing fully-charged batteries to

EV owners, but also can capture the variability in solar PV generation associated with the prosumers. By doing this, the power needed to be injected to the feeder by the utility grid can be controlled to some extent. Fig. 1 shows the BSS-based model architecture for capturing distribution grid-integrated solar variability, where the power of $P_t^u = P_t^M + \sum_{j \in N} P_{jt}^c$ is provided by the utility grid to this distribution feeder. Nevertheless, the BSS is expected to receive incentive from the utility grid to capture the variability of solar generation.

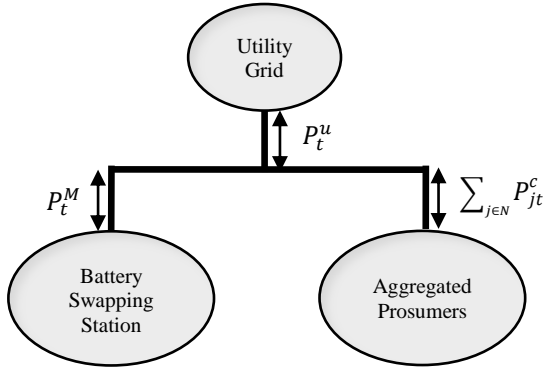

Fig. 1. BSS-based architecture for capturing distribution grid-integrated solar variability.

A model for the BSS optimal scheduling problem is proposed from the BSS owner's perspective. The objective of the proposed model is to minimize the BSS total operation cost, which represents the accumulated cost of exchanging energy with the utility grid, while taking into account the output power adjustment for capturing solar generation variability. The proposed model is subject to four sets of constraints associated with the utility grid, the BSS, individual batteries, and solar mitigation. A mixed-integer programing method is utilized to formulate the BSS optimal scheduling model from the BSS owner's view.

## III. PROBLEM FORMULATION

The BSS owner's objective is to minimize its operation cost, i.e., the cost of exchanging power with the utility grid, as in (1). The quantity of power exchange with the utility grid is determined by subtracting the accumulated battery charging power from the discharging power as in (2). This quantity can be positive or negative as for power import or export, respectively.

$$\min \sum_t [P_t^M \rho_t \tau] \qquad (1)$$

$$P_t^M = \sum_b (P_{bt}^{ch} - P_{bt}^{dch}) \qquad \forall t \qquad (2)$$

Based on the forecasted hourly electricity price $\rho$, the operation cost is calculated. As the power exchange with the utility grid can be positive or negative, the objective function can be positive or negative which means the BSS owner not only is able to minimize its cost, but can also make revenue through exporting power to the utility grid. $\tau$ denotes time period, which can be set according to the BSS owner's discretion. By considering shorter time periods, the BSS can more accurately capture the rapid variability of solar generation. However, the proper choice of the time period is a tradeoff between the accuracy and the computation time. The objective function of the proposed model is subject to the following constraints.

### A. Grid Constraints

The sum of transferred power for charging/discharging batteries in each time period is limited by the flow limits of the line connecting the BSS to the utility grid, as represented in (3).

$$-P^{M,\max} \leq P_t^M \leq P^{M,\max} \qquad \forall t \qquad (3)$$

### B. BSS System Constraints

The BSS constraints are employed to model available fully-charged batteries in order to meet the battery swapping demand in each time period, as formulated in (4)-(5).

$$1 + M(C_{bt} - C^{\max}) \leq x_{bt}^F \leq 1 + \varepsilon(C_{bt} - C^{\max}) \quad \forall t, \forall b \qquad (4)$$

$$D_t = \sum_b x_{b(t-1)}^F \qquad \forall t \qquad (5)$$

To determine whether the battery is fully-charged or not, (4) is proposed. If $C_{bt}$ is equal to $C_b^{max}$, battery $b$ is fully-charged and binary variable $x_{bt}^F$ is set to one, which indicates battery $b$ is ready to be swapped in the next time period. Otherwise, if $C_{bt}$ less than $C_b^{max}$, the battery is not fully-charged and the binary variable $x_{bt}^F$ is forced to be 0, which means battery $b$ is not ready for swapping. The balance equation (5) ensures that the number of the fully-charged batteries in the previous time period is equal to the current swapping demand. In other words, once a battery is fully-charged, it will be swapped in the subsequent time period.

### C. Individual Battery and Charger Constraints

The battery and charger constraints are directly resulted from their technologies and include limitations associated with power rating and stored energy. These constraints are defined to ensure that the batteries and chargers do not exceed their associated operational limits.

$$0 \leq P_{bt}^{ch} \leq P_b^{ch,\max} \qquad \forall t, \forall b \qquad (6)$$

$$0 \leq P_{bt}^{dch} \leq P_b^{dch,\max} \qquad \forall t, \forall b \qquad (7)$$

$$C^{\min} \leq C_{bt} \leq C^{\max} \qquad \forall t, \forall b \qquad (8)$$

$$-M(1 - x_{b(t-1)}^F) \leq C_{bt} - C_{bt}^{ini} - \eta^{ch} P_{bt}^{ch} \tau + \eta^{dch} P_{bt}^{dch} \tau \leq$$
$$\leq M(1 - x_{b(t-1)}^F) \qquad \forall t, \forall b \qquad (9)$$

$$-M(x_{b(t-1)}^F) \leq C_{bt} - C_{b(t-1)} - \eta^{ch} P_{bt}^{ch} \tau + \eta^{dch} P_{bt}^{dch} \tau \leq$$
$$\leq M(x_{b(t-1)}^F) \qquad \forall t, \forall b \qquad (10)$$

Charging/discharging power rating of each battery is limited by the maximum charging/discharging power which are assumed to be positive (6)-(7). Equation (8) ensures that the batteries are operating within their associated capacity limits. Based on (8), the battery stored energy is limited by its maximum and minimum limits. Equations (9)-(10) are defined to calculate the battery stored energy according to the value of charged/discharged power and charging/discharging efficiency. When a battery is fully-charged in the previous time period (i.e., $x_{b(t-1)}^F=1$), it will be swapped with an empty battery in the next time period, and consequently this empty battery with the initial stored energy of $C_{bt}^{ini}$ is charged/discharged based on (9). Without the loss of generality, when a battery is not fully-charged in the previous time period (i.e., $x_{b(t-1)}^F=0$), it is charged/discharged based on (10).

### D. Solar Variability Constraints

The solar variability constraints are introduced to capture the variability caused by solar generation. The BSS exchanged power with the utility grid is utilized for mitigating PV output fluctuations.

$$-\Delta^u - \Delta_t \le P_t^M - P_{(t-1)}^M \le \Delta^u - \Delta_t \qquad \forall t \qquad (11)$$

$$\Delta_t = \sum_j P_{jt}^c - \sum_j P_{j(t-1)}^c \qquad \forall t \qquad (12)$$

Equation (11) is defined to capture the aggregated prosumers net loads variability, where $\Delta^u$ denotes the amount of variability being captured by the utility grid, and the rest is picked up by the BSS. The aggregated prosumers net loads variability between two successive time periods (i.e., $\Delta_t$) is formulated in (12). Nevertheless, leveraging (11) and (12), the aggregated prosumers net load variability is entirely captured by the BSS through exchanged power with the utility grid.

### E. Uncertainty Consideration

To capture variability of solar generation, the proposed model uses hourly forecasted values of solar generation. As the solar generation is affected by weather conditions which are uncontrollable, the forecasting errors are inevitable. To deal with the solar generation uncertainty, a robust optimization method will be utilized. By maximizing the minimum value of the objective (1) over a defined uncertainty set, i.e., solar generation uncertainty, the worst-case solution will be determined. The uncertain parameter, i.e., solar generation, is assumed to be within an interval around the forecasted value, i.e., a polyhedral uncertainty set. By increasing maximum number of instances that this uncertain parameter can differ from its forecasted value, which is called the budget of uncertainty, the robustness of the solution will increase, while reducing the solution optimality.

## IV. NUMERICAL SIMULATIONS

The performance of the proposed model is analyzed with a BSS consisting of 300 batteries with the individual capacity of 100 kWh. The BSS is equipped with 300 AC-level-2 battery chargers with the maximum power of 17.2 kW for a 100-kWh configured battery [31]. It is assumed that there is no power transfer limit between the BSS and the utility grid. The time period is set to be 1 h, i.e, $\tau=1$ h, where the proposed model for BSS optimal scheduling model is solved for a 24-h horizon. The maximum value of variability desired to be captured by the utility grid, i.e., $\Delta^u$, is assumed to be 1 MW/h. It means that the BSS is used to capture the aggregated prosumers net loads variability above this value.

The day-ahead forecasted values of electricity price over the 24-h horizon are given in Table I. The aggregated load data, solar generation, and consequently the net load for a sample distribution feeder are listed in Table II. The BSS demand over the 24-h horizon is tabulated in Table III. The proposed BSS optimal scheduling problem is solved using CPLEX 11.0 by a personal computer with Intel Core i5, 2.3 GHz processor, and 4 GB RAM. The computation time for each of the following cases is less than 10 min, which advocates the computational efficiency of the proposed model. The following cases are studied:

**Case 1**: BSS optimal scheduling ignoring solar variability constraints.
**Case 2**: BSS optimal scheduling with solar variability constraints.
**Case 3**: BSS optimal scheduling under solar generation uncertainty.

TABLE I
HOURLY ELECTRICITY PRICE

| Time (h) | 1 | 2 | 3 | 4 | 5 | 6 |
|---|---|---|---|---|---|---|
| Price ($/MWh) | 15.03 | 10.97 | 13.51 | 15.36 | 18.51 | 21.8 |
| Time (h) | 7 | 8 | 9 | 10 | 11 | 12 |
| Price ($/MWh) | 17.3 | 22.83 | 21.84 | 27.09 | 37.06 | 68.95 |
| Time (h) | 13 | 14 | 15 | 16 | 17 | 18 |
| Price ($/MWh) | 65.79 | 66.57 | 65.44 | 79.79 | 115.45 | 110.28 |
| Time (h) | 19 | 20 | 21 | 22 | 23 | 24 |
| Price ($/MWh) | 96.05 | 90.53 | 77.38 | 70.95 | 59.42 | 56.68 |

TABLE II
HOURLY BSS DEMAND

| Time (h) | 1 | 2 | 3 | 4 | 5 | 6 |
|---|---|---|---|---|---|---|
| Demand (No.) | 2 | 1 | 1 | 2 | 4 | 6 |
| Time (h) | 7 | 8 | 9 | 10 | 11 | 12 |
| Demand (No.) | 8 | 7 | 6 | 5 | 5 | 4 |
| Time (h) | 13 | 14 | 15 | 16 | 17 | 18 |
| Demand (No.) | 6 | 7 | 8 | 10 | 12 | 12 |
| Time (h) | 19 | 20 | 21 | 22 | 23 | 24 |
| Demand (No.) | 9 | 8 | 6 | 5 | 2 | 1 |

TABLE III
HOURLY AGGREGATED PROSUMERS SOLAR GENERATION, LOAD, AND NET LOAD IN A DISTRIBUTION FEEDER

| Time (h) | 1 | 2 | 3 | 4 | 5 | 6 |
|---|---|---|---|---|---|---|
| Solar (MW) | 0 | 0 | 0 | 0 | 0 | 0 |
| Load (MW) | 6.75 | 6.25 | 5.90 | 5.85 | 6.05 | 6.25 |
| Net Load (MW) | 6.75 | 6.25 | 5.90 | 5.85 | 6.05 | 6.25 |
| Time (h) | 7 | 8 | 9 | 10 | 11 | 12 |
| Solar (MW) | 0 | 0 | 0.50 | 2.00 | 4.00 | 5.75 |
| Load (MW) | 6.40 | 7.00 | 7.30 | 7.60 | 8.00 | 8.50 |
| Net Load (MW) | 6.40 | 7.00 | 6.80 | 5.60 | 4.00 | 2.75 |
| Time (h) | 13 | 14 | 15 | 16 | 17 | 18 |
| Solar (MW) | 7.00 | 7.10 | 7.00 | 6.20 | 5.50 | 3.00 |
| Load (MW) | 9.25 | 9.00 | 8.50 | 8.35 | 8.50 | 9.00 |
| Net Load (MW) | 2.25 | 1.90 | 1.50 | 2.15 | 3.00 | 6.00 |
| Time (h) | 19 | 20 | 21 | 22 | 23 | 24 |
| Solar (MW) | 1.35 | 0.40 | 0 | 0 | 0 | 0 |
| Load (MW) | 10.15 | 10.35 | 9.50 | 8.50 | 7.25 | 6.90 |
| Net Load (MW) | 8.80 | 9.95 | 9.50 | 8.50 | 7.25 | 6.90 |

**Case 1:** The BSS optimal scheduling is studied while ignoring the solar variability constraints. It means that the BSS in the distribution feeder does not participate in capturing the aggregated prosumers net load variability. As the BSS optimal schedule in this case aims at minimizing its operation cost, i.e., focusing on the BSS price-based scheduling, it is expected that the utility grid experiences a severe net load variability. The BSS operation cost is calculated as $-1555.72 in this case. This negative value for the operation cost means that the BSS owner makes money through energy arbitrage. Fig. 2 shows the distribution feeder net load ($P_{u,t}$) and the BSS exchanged power with the utility grid ($P_{M,t}$).

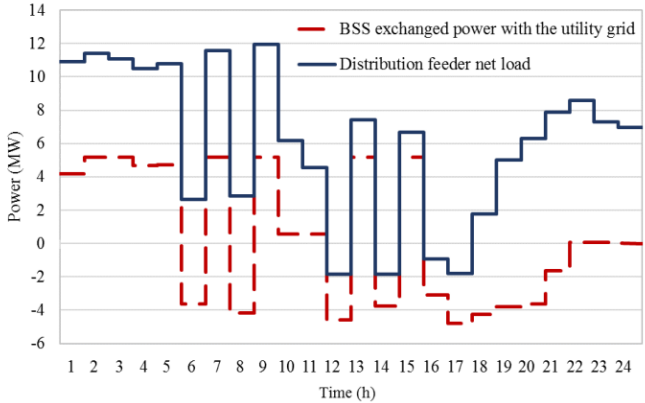

Fig. 2. BSS exchanged power with the utility grid and distribution feeder net load in Case 1.

The trend of power exchange with the utility grid in this case is based on energy arbitrage revenue, which is purchasing electricity at low price hours for charging batteries and selling back electricity through discharging at high price hours. Moreover, since the solar variability constraints are ignored, the BSS price-based schedule is targeted to minimize its operation cost, so that the utility grid undergoes a severe net load variability in the distribution feeder. For instance, there are the severe net load changes of 9.11 MW/h and 9.24 MW/h between hours 9-10 and 13-14, respectively, which must be captured by the utility grid.

**Case 2:** The BSS optimal scheduling is studied considering the solar variability constraints. In this case, the BSS exchanged power with the utility grid contributes in capturing the aggregated prosumers net load variability. As the utility grid is to capture the aggregated prosumers net loads variability of less than 1 MW/h, any variability larger than this value is captured by the BSS based on the proposed model. Compared to Case 1, the BSS operation cost is increased by 21.8% to $-1216.14, which translates into less benefits for the BSS owner. Nevertheless, the BSS exchanged power with the utility grid is reshaped in such a way that the distributed solar generation variability is captured at the expense of increased operation cost for the BSS.

As the BSS operation cost is increased, the grid operator not only should pay to the BSS owner to compensate this increase, but also should incentivize the BSS owner to contribute in mitigating the solar generation variability as well as helping the power systems for hosting a higher penetration of solar generation. Fig. 3 compares the distribution feeder net load with and without variability constraints.

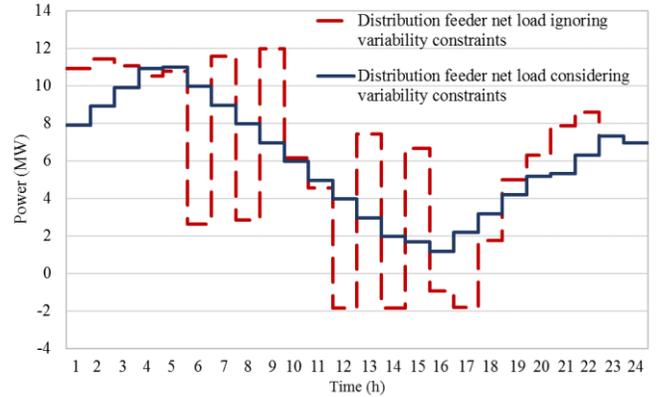

Fig. 3. Comparison of distribution feeder net load in Case 1 and 2.

As shown in Fig. 3, the maximum changes in distribution feeder net load supplied by the utility grid is bounded to be less than 1 MW/h, which makes the distribution feeder net load smoother.

**Case 3:** In this case, the BSS optimal scheduling problem with solar variability constraints is studied under solar generation uncertainty. Accordingly, the BSS owner's objective (1) is maximized over solar generation uncertainty to achieve the worst-case solution using a robust optimization approach. Solar generation forecast error is considered to be ±20%. The sensitivity of the BSS operation cost with respect to the uncertainty budget is carried out, where the obtained results are listed in Table IV.

TABLE IV
BSS OPERATION COST WITH RESPECT TO UNCERTAINTY BUDGET

| | Uncertainty Budget (hours/day) | | | | |
|---|---|---|---|---|---|
| | 0 | 3 | 6 | 9 | 12 |
| BSS Operation Cost ($) | -1216.14 | -1108.35 | -967.1 | -921.07 | -910.49 |

The obtained results advocate the fact that by increasing the uncertainty budget, the BSS operation cost increases, which translates into a reduction in the BSS owner's benefits. This increase in the BSS operation cost indicates the cost of robustness which are paid to make the BSS operation more robust against solar generation uncertainty. Nevertheless, this study demonstrates that the BSS owner achieves an optimal scheduling at higher cost when the forecasted data are uncertain.

## V. CONCLUSIONS

This paper introduced the BSS as an energy storage to address solar generation variability in distribution networks. A BSS optimal scheduling model was proposed from the BSS owner's perspective with the objective of capturing distribution grid-integrated solar variability. To this end, the BSS exchanged power with the utility grid was reshaped in such a way that the distributed solar generation variability was captured. Using mixed-integer linear programming, the proposed model was formulated to minimize the BSS operation cost, while taking into account the prevailing constraints associated with the utility grid power exchange, the

BSS system, individual batteries, and solar variability. The proposed model was investigated through numerical simulations, where it was demonstrated that the BSS provides a viable approach in capturing the solar generation variability as well as helping the utility grids for hosting a higher penetration of solar generation.